\RequirePackage{lineno}
\documentclass{nature}
\usepackage{xcolor}
\usepackage[T1]{fontenc}
\usepackage{caption}
\usepackage{amsmath}
\usepackage{amssymb}
\usepackage{amsfonts}
\usepackage{bm}
\usepackage{textcomp}
\usepackage[colorlinks]{hyperref}
\usepackage{graphicx}
\usepackage{dcolumn}
\usepackage{float}
\usepackage{cleveref}

\makeatletter
\let\saved@includegraphics\includegraphics
\AtBeginDocument{\let\includegraphics\saved@includegraphics}
\makeatother

\hypersetup{
citecolor={blue},
urlcolor={blue}
}

\crefname{appendix}{Appendix}{Appendices}
\crefname{equation}{Eq.}{Eqs.}
\crefname{figure}{Fig.}{Figs.}
\crefname{table}{Table}{Tables}
\crefname{section}{Section}{Sections}
\crefname{mythe}{Theorem}{Theorems}
\crefname{mydef}{Definition}{Definitions}
\makeatletter
\newcommand{\dummylabel}[2]{\def\@currentlabel{#2}\label{#1}}
\makeatother

\title{First-principles Floquet analysis from real-time propagation}

\author{Ruipeng Li$^{1,\dagger}$, Benshu Fan$^{1,\dagger}$, Umberto De Giovannini$^{1,2}$, Hannes Hübener$^{1,*}$, Angel Rubio$^{1,3,\ddagger}$}

\begin{document}
\maketitle

\begin{affiliations}
\item Max Planck Institute for the Structure and Dynamics of Matter and Center for Free-Electron Laser Science, Luruper Chaussee 149, 22761, Hamburg, Germany
\item Dipartimento di Fisica e Chimica-Emilio Segr\`e, Universit\`a degli Studi di Palermo, Via Archirafi 36, 90123 Palermo, Italy
\item Initiative for Computational Catalysis, The Flatiron Institute, 162 Fifth Avenue, New York, NY 10010, United States\\

$\dagger$ These authors contributed equally to this work.\\
$*$ hannes.huebener@mpsd.mpg.de\\
$\ddagger$ angel.rubio@mpsd.mpg.de
\end{affiliations}

\section*{Abstract}
\begin{abstract}
We present a real-time Floquet analysis method for extracting quasi-energies and Floquet states directly from propagated wavefunctions. By reconstructing the one-period evolution operator from overlaps between time-evolved states, the method avoids the explicit construction of the enlarged Floquet Hamiltonian and adds negligible computational overhead to time-dependent simulations. To resolve the ambiguity inherent in the reduced-zone representation, we introduce an unfolding procedure based on the harmonic decomposition of Floquet states, which recovers their underlying equilibrium band character beyond the reduced Floquet Brillouin zone. The reconstructed wavefunctions further provide access to the symmetry properties of individual light-induced sidebands. We demonstrate the applicability and generality of the method in first-principles time-dependent simulations of real materials, ranging from two-dimensional monolayers to a three-dimensional bulk semiconductor, and including finite pulses without strict time periodicity. This framework directly connects real-time simulations with Floquet observables, enabling practical analysis of light-driven electronic structure in materials.
\end{abstract}

\section*{Introduction}
The control of material properties through periodic driving has emerged as a central paradigm in condensed matter physics\cite{Hsieh:2017ix,DelaTorre2021,bloch2022strongly,bao2022light}. Within this framework, Floquet theory provides a description of quantum systems subject to time-periodic fields in terms of quasi-energies and time-periodic states\cite{Sambe:1973hi,oka2019floquet}. This perspective has enabled proposals for engineering electronic properties such as light-induced sidebands\cite{Aeschlimann2021Survival,merboldt2025observation,choi2025observation}, band renormalization\cite{deGiovannini:2016cb,Claassen:2016ge,Ofer2022PRR,zhou2023pseudospin,zhou2023Floquet}, gap opening\cite{bao2024light,fei2026observation}, and topological phases\cite{oka_photovoltaic_2009,lindner_floquet_2011,Kitagawa:2011dr,Wang:2013fe,Sentef:2015jp,sie2015valley,Mahmood:2016bu,chan2016chiral,Hubener:2017ht,sie2017large,Sato2019a,Lindner2020,schuler2020circular,mciver2020light,ito2023build,liu2023floquet,zhu2023floquet,fanbs2024chiral,Rubio2024JPCM,zhan2024perspective,li2025light,fragkos2025floquet}.

Despite its conceptual power, the practical application of Floquet theory to realistic materials remains challenging\cite{Caruso_2026}. The explicit construction of the Floquet Hamiltonian requires an enlarged Hilbert space combining electronic states and Floquet harmonic sectors, leading to substantial computational overhead\cite{Giovannini_2020}. In addition, in first-principles simulations based on time-dependent density functional theory (TDDFT), the natural output is the real-time propagation of electronic states rather than Floquet eigenstates themselves. These aspects complicate the direct extraction and interpretation of Floquet observables in realistic non-equilibrium settings. A further difficulty is that the formalism relies on strict time periodicity, whereas experimentally relevant situations often involve finite pulses, transient regimes, and incomplete pump--probe overlap. Nevertheless, a growing number of studies have shown that Floquet concepts remain highly relevant beyond strictly periodic conditions\cite{Lucchini.2022,chan2023giant,xiao2025interaction,pareek2026driving}. In particular, recent high-profile pump--probe time-resolved photoemission experiments have shown that light-dressed electronic
structures can be naturally interpreted in terms of transient Floquet quasi-energy bands\cite{zhou2023Floquet,zhou2023pseudospin,merboldt2025observation,choi2025observation,fei2026observation}. This suggests that Floquet theory is not only a computational recipe for constructing periodically driven steady states, but also a powerful analysis framework for non-equilibrium electronic structure and, more generally, transient quasiparticle dynamics. An important open question is therefore how Floquet properties can be efficiently extracted and interpreted directly from real-time simulations.

In this work, we develop a real-time Floquet analysis framework for first-principles simulations that extracts Floquet quasi-energies and Floquet states directly from propagated wavefunctions. The key step of our approach is to reconstruct a representation of the evolution operator from overlaps between time-evolved states separated by one driving period. Diagonalizing this operator yields the Floquet quasi-energies and eigenvectors within the propagated-state subspace. A different scheme for obtaining Floquet properties directly from the time-dependent wavefunctions has previously been described in Ref.~\citenum{Kapoor:2012jp}, however without making use of the one-period evolution operator, which is central to the simplicity of our approach. Additionally, to make the resulting spectra physically interpretable, we introduce an unfolding scheme based on the harmonic decomposition of Floquet states. Since Floquet quasi-energies are defined modulo the driving frequency, physically relevant band features can be obscured in the reduced Floquet Brillouin zone (FBZ) by folded photon replicas. The zeroth-harmonic weights, evaluated in the appropriate extended-zone representation, identify which branches retain the character of the underlying equilibrium bands, allowing band edges, light-induced replicas, and hybridization gaps to be resolved. Beyond quasi-energy spectra, the reconstructed real-space Floquet wavefunctions enable the symmetry character of individual light-induced sidebands to be analyzed. We demonstrate the approach in first-principles real-time TDDFT simulations of driven materials, spanning two-dimensional (2D) monolayers and a fully periodic three-dimensional (3D) bulk solid. We further show that Floquet-like electronic structures can be meaningfully identified in finite-pulse regimes, providing a practical framework for analyzing transient non-equilibrium phenomena beyond the strictly periodic limit.

\section*{Results}
\label{sec:results}

\subsection{Floquet states and time-evolution operator}

We consider a quantum system governed by a time-periodic Hamiltonian $H(t+T)=H(t)$, with period $T = 2\pi/\Omega$, where $\Omega$ is the 
frequency of the drive. Throughout, we set the reduced Planck constant $\hbar=1$, so that $\Omega$ also 
denotes the corresponding photon energy. According to Floquet theory, all solutions of the time-dependent Schr\"{o}dinger equation can be written as a linear combination of Floquet states defined as
\begin{equation}\label{eq:F_def}
|\Psi_\alpha(t)\rangle = e^{-i E_\alpha t} |\phi_\alpha(t)\rangle,
\end{equation}
where $|\phi_\alpha(t)\rangle = |\phi_\alpha(t+T)\rangle$ are time-periodic functions and $E_\alpha$ are the quasi-energies. The periodic functions can further be expanded in a Fourier series 
\begin{equation}\label{eq:phi_def}
 |\phi_\alpha(t)\rangle = \sum_n e^{i n \Omega t} |u^{(n)}_\alpha \rangle,
\end{equation}
where $|u_\alpha^{(n)}\rangle$ are the static components that take the role of eigenstates corresponding to the energies $E_\alpha$. This can be seen by inserting Eq.~(\ref{eq:F_def}) and (\ref{eq:phi_def}) into the time-dependent Schr\"odinger equation and performing the inverse Fourier series expansion of Eq.~(\ref{eq:phi_def}):
\begin{equation}\label{eq:HF}
\sum_m\left(\int_T dt e^{i (m-n) \Omega t} H(t) +m\Omega\delta_{mn}\right)|u_\alpha^{(m)}\rangle = E_\alpha |u_\alpha^{(n)}\rangle,
\end{equation}
where $\int_Tdt$ denotes the weighted average integral over one period. This has the structure of a static eigenvalue equation $\mathcal{H}u_\alpha=E_\alpha u_\alpha$, where $\mathcal{H}$ is referred to as Floquet Hamiltonian, defined by Eq.~(\ref{eq:HF}). The $u_\alpha$ are the eigenvectors, with components corresponding to the components in the Fourier series Eq.~(\ref{eq:phi_def}). Therefore, diagonalization of Eq.~(\ref{eq:HF}) gives the ingredients for the construction of the Floquet states. This formulation has, however, some drawbacks. One first needs to construct the Floquet Hamiltonian, which has a significantly enlarged Hilbert space with respect to the static or time-dependent Hamiltonian. Although the Floquet Hamiltonian can be constructed in certain situations within a restricted electronic subspace containing only states relevant to the energy window of interest, each retained electronic state must still be replicated over the included Floquet sectors. In practice one has to choose a maximum number $n_\textrm{max}$ of Floquet components such that within the energy range of interest in the band structure the eigenvalues are independent of the $n_\textrm{max}$. Since the eigenvalues of the Floquet Hamiltonian are strictly periodic in energy, $E_{\alpha'}=E_{\alpha}+n\Omega$, this can require a large number of essentially redundant bands, and by extension, redundant eigenvectors. The resulting Floquet Hamiltonian therefore has a dimension that grows with both the number of electronic states included in the chosen subspace and the number of retained Floquet sectors.

Instead, an equivalent formulation of Floquet theory can be obtained in terms of the time-evolution operator, which avoids this additional enlargement associated with the Floquet-sector dimension (cf. for example Ref.~\citenum{rudner2020floquet}). The evolution of one period $T$ of a Floquet state in Eq.~(\ref{eq:F_def}), $U(t+T,t)|\Psi_\alpha(t)\rangle=|\Psi_\alpha(t+T)\rangle$, reads
\begin{equation}\label{eq:U_eig}
 U(t+T,t)|\phi_\alpha(t)\rangle = e^{-iE_\alpha T}|\phi_\alpha(t)\rangle=\mathcal{E}_\alpha|\phi_\alpha(t)\rangle,
\end{equation}
where we have used the time-periodicity of $|\phi_\alpha(t)\rangle$. We note that the full Floquet states are also eigenstates of the one-period time evolution operator with the same eigenvalues. Both states differ up to a complex phase and hence it is a matter of choice which states are used. The above equation shows that $|\phi_\alpha(t)\rangle$ are the eigenstates of the one-period time-evolution operator and the Floquet quasi-energies can be obtained from $E_\alpha=-\arg(\mathcal{E}_\alpha)/T$. Thus, diagonalizing $U(t+T,t)$ provides direct access to the quasi-energies and Floquet states from the evolution operator itself. In the next section we will describe how this formulation can be applied in practice for the analysis of pump-probe excitations.

\subsection{Construction of the evolution operator from real-time propagation}
\noindent
Here, we use this established evolution-operator formulation as the basis of a computationally efficient analysis framework for first-principles real-time simulations. The dimension of the reconstructed evolution operator is determined only by the number of propagated states and does not acquire the additional Floquet-sector dimension of the conventional Floquet Hamiltonian. Its construction and diagonalization therefore constitute a low-cost post-processing step on top of the real-time propagation. The distinctive aspects of the present approach are the subsequent reconstruction of harmonic-resolved Floquet wavefunctions, together with the associated Floquet-Brillouin-zone unfolding and real-space symmetry analysis of individual light-induced sidebands. In real-time approaches such as TDDFT, the fundamental quantity is the propagated state
\begin{equation}
|\psi_i(t)\rangle = U(t,0)|\psi_i(0)\rangle
\end{equation}
obtained by numerical integration of the time-dependent Schr\"{o}dinger equation. The key observation underlying the present method is that the action of the evolution operator is encoded in these propagated states and we can thus use them to reconstruct a representation of this operator.

Let $\{ |\psi_i(t)\rangle \}$ denote a set of orthonormal states at time $t$, for example the time-evolved Kohn--Sham (KS) orbitals. The matrix elements of the one-period evolution operator in this basis can be constructed from overlaps of time-evolved states. Propagation over one period from time $t$ to $t+T$ represents the following action of the time evolution operator
\begin{equation}
  | \psi_j(t+T) \rangle = U(t+T,t) | \psi_j(t) \rangle.
\end{equation}
From this follows
\begin{equation}
\langle \psi_i(t) | \psi_j(t+T) \rangle=\langle \psi_i(t) | U(t+T,t) | \psi_j(t) \rangle = U_{ij}(t+T,t).
\end{equation}
This means the matrix representation of the one-period evolution operator can be obtained directly from the time-evolving states with those at time one period back in time. The resulting matrix has the dimension of the number of propagated states. This representation is exact within the propagated-state subspace. The quasi-energies and eigenstates in the energy range of interest must be converged with respect to the number of states used.

In the real-time TDDFT calculations considered here, the time-dependent KS Hamiltonian is
\begin{equation}\label{eq:tdks_selfconsistent}
H_{\rm KS}[n](t)=h_0(t)+v_{\rm H}[n](t)+v_{\rm xc}[n](t),
\end{equation}
where $n(\mathbf{r},t)$ is the time-dependent electronic density, 
$v_{\rm H}[n](t)$ and $v_{\rm xc}[n](t)$ are the Hartree and 
exchange--correlation potentials, respectively, and $h_0(t)$ contains the kinetic, ionic, and explicitly time-dependent laser-coupling terms. The density, Hartree potential, and exchange-correlation potential are updated self-consistently at every propagation step. Thus, the overlap matrix is constructed from the evolution generated by the full self-consistent KS Hamiltonian, without a frozen-potential approximation. For the adiabatic local-density approximation employed here, $v_{\rm H}$ and $v_{\rm xc}$ are instantaneous functionals of the density. If $n(\mathbf r,t+T)=n(\mathbf r,t)$, they inherit the same periodicity and the reconstructed operator is the one-period propagator of a self-consistent periodic KS Hamiltonian. Their dynamical contributions need not be negligible: they may modify screening, level positions, and light--matter coupling while preserving the Floquet construction. Conversely, periodicity of the external field alone does not guarantee periodicity of the self-consistent KS Hamiltonian. Population of several dressed states, carrier redistribution, or collective density dynamics can introduce additional timescales into the density and the Hartree--exchange--correlation potential\cite{Kapoor2013Periodicity}. In that regime, the operator reconstructed over one optical cycle has a time-local rather than a stationary Floquet interpretation. For the cases considered in our paper, dynamical $v_{\rm Hxc}$ effects are expected to be weak, and one obtains qualitatively similar results even with a static $v_{\rm Hxc}$ potential. However, our full propagation approach, which explicitly accounts for the dynamics of $v_{\rm Hxc}$, is more general and captures regimes involving the coherent excitation of internal material modes, such as Floquet excitons\cite{chan2023giant}.

In a TDDFT propagation using an instantaneous density functional, the periodicity of the time-dependent density can be assessed directly from the cycle-to-cycle density deviation
\begin{equation}\label{eq:density_periodicity}
\delta n(t)=\left[\frac{\displaystyle\int_t^{t+T}d\tau\,\left\|n(\tau+T)-n(\tau)\right\|_2^2}{\displaystyle\int_t^{t+T}d\tau\,\left\|n(\tau)\right\|_2^2}\right]^{1/2},
\end{equation}
where the spatial $L^2$ norm $\|\cdot\|_2$ is taken over the simulation cell. Establishing the applicability of the Floquet interpretation requires verifying that $\delta n(t)\ll1$, together with a slowly varying external-field envelope, over the chosen analysis window. Equivalently, one may monitor the cycle-to-cycle variation of the full KS potential. If substantial carrier excitation, energy absorption, screening dynamics, or collective motion produces an appreciable cycle-to-cycle variation, the eigenvalues of $U(t+T,t)$ should instead be regarded as local, time-dependent spectral quantities, and the stationary Floquet interpretation becomes less well defined.

For numerical time propagation schemes, both the projection and subsequent diagonalization of the matrix come at negligible additional computational cost and hence provide a very attractive alternative to the construction of the Floquet Hamiltonian. Importantly, the evolution operator $U(t+T,t)$ can be constructed for any starting time $t$ during the propagation. Its interpretation as a stationary or time-local Floquet operator is, however, subject to the periodicity conditions discussed above. In order to put this to practical use one still has to face the issue of FBZ unfolding, which we discuss in the next section.

\subsection{Floquet band unfolding and quasiparticle character}
\label{sec:weight}

The quasi-energies obtained from the eigenvalues of the time-evolution operator are defined modulo $\Omega$, reflecting the periodicity of the Floquet spectrum in energy space. As a consequence, the spectrum is restricted to a FBZ, typically defined as $E_\alpha \in [-\Omega/2, \Omega/2]$. While this reduced-zone representation is formally complete, it complicates the interpretation of electronic band structures in solids. All photon replicas of different bands are folded onto the same energy interval and we will show below (cf. Fig.~\ref{fig:floquet_graphene}) that it becomes difficult to identify physically relevant features such as band edges or hybridization gaps. Simply repeating the FBZ does not resolve this ambiguity, as the correspondence between replicas and underlying equilibrium bands is generally non-trivial. Therefore we present here an approach for unfolding the FBZ based on the harmonic components of Floquet eigenstates. The following is the main technical development reported in this paper.

The key observation is that the harmonic components $|u^{(n)}_\alpha\rangle$ of a Floquet eigenstate contain direct information about the physical character of the corresponding Floquet band $E_\alpha$. The index $n$ labels the component in the harmonic expansion. In the adiabatic limit, where the driving is sufficiently slow and weak, the Floquet state remains almost identical to the equilibrium electronic state. In this case only the $n=0$ component is appreciable, while all higher harmonic components vanish. Away from the adiabatic regime, spectral weight is then redistributed among different harmonic components, reflecting the increasing influence of the periodic drive on each state\cite{Cai2026Occupation}. In the extreme case of resonant hybridization between an equilibrium band and a replica, two harmonic components contribute with comparable weight, signaling strong mixing between the states. This observation allows us to identify the equilibrium character of a Floquet state through the weight of its zeroth harmonic component. We therefore define
\begin{equation}\label{eq:weight}
w_\alpha  = |u_\alpha^{(0)}|,
\end{equation}
that is, the norm of the $n=0$ Fourier component of the Floquet eigenstate. This quantity measures how strongly a Floquet state retains the character of the original equilibrium band structure, and therefore provides a natural way to distinguish physically relevant bands from higher-order Floquet replicas. 

We use $w_\alpha$ as a weighting factor to analyze and visualize the Floquet band structure beyond the FBZ. When quasi-energy spectra are unfolded by adding integer multiples of $\Omega$, the zeroth-harmonic weight allows one to identify the physically relevant branches corresponding to the underlying electronic bands. States dominated by higher harmonics carry small $w_\alpha$ and are systematically suppressed, while states with large $w_\alpha$ retain the character of the original bands. However, one should be careful that, beyond the FBZ, the zeroth-order harmonic component is no longer identical to $|u_\alpha^{(0)}\rangle$ as defined above. In fact, when we shift the Floquet energy by integer multiples of driving frequency, namely $E_{\alpha'}=E_\alpha-r\Omega$, the corresponding harmonic components should also be shifted with the same integer $r$ in its index, as a result of the Floquet gauge redundancy (see Methods for a detailed discussion). This harmonic decomposition of Floquet eigenstates defines a physically meaningful notion of equilibrium band character for non-equilibrium quasiparticles.

To evaluate the weights $w_\alpha$ we need a further computational step. In contrast to the eigenvalue problem of the Floquet Hamiltonian in Eq.~(\ref{eq:HF}), which directly yields the $|u_\alpha^{(n)}\rangle$ vectors, diagonalization of the matrix $[U_{ij}]$ gives the periodic parts of the Floquet states, $|\phi_\alpha(t)\rangle$. To construct $w_\alpha$ we thus need to find the components of the expansion Eq.~(\ref{eq:phi_def}):
\begin{equation}\label{eq:un}
    |u^{(n)}_{\alpha}\rangle = \int_T dt e^{-i n \Omega t}|\phi_\alpha(t)\rangle.
\end{equation}
The periodic Floquet state $|\phi_\alpha(t)\rangle$ can be constructed at arbitrary times by considering $|\Psi_\alpha (t)\rangle=U(t,t_0)|\Psi_\alpha (t_0)\rangle= U(t,t_0) e^{-iE_\alpha t_0} |\phi_\alpha(t_0)\rangle$ giving
\begin{equation}\label{eq:phit1}
    |\phi_\alpha(t)\rangle = U(t,t_0) e^{-iE_\alpha (t_0-t)} |\phi_\alpha(t_0)\rangle.
\end{equation}
Taking $t_0$ as the starting time for the Floquet analysis, i.e. the time for which the eigenvalue problem Eq.~(\ref{eq:U_eig}) was solved, we have the expansion $|\phi_\alpha(t_0)\rangle = \sum_i c_{\alpha i} |\psi_i(t_0)\rangle$, where $c_{\alpha i}$ are the components of eigenvectors of $[U_{ij}]$. Inserting this expansion into Eq.~(\ref{eq:phit1}) gives
\begin{eqnarray}
    |\phi_\alpha(t)\rangle &=& e^{-iE_\alpha (t_0-t)} \sum_i  c_{\alpha i} U(t,t_0) |\psi_i(t_0)\rangle \\ 
    &=& e^{-iE_\alpha (t_0-t)} \sum_i  c_{\alpha i} | \psi_i(t)\rangle,\label{eq:phit}
\end{eqnarray}
where $|\psi_i(t)\rangle$ are still the time-evolving states solved by the computational time propagation. From Eq.~(\ref{eq:phit}) one can now construct the static Floquet components in Eq.~(\ref{eq:un}) by explicitly evaluating the time-integral:
\begin{equation}\label{eq:u_alpha}
    |u^{(n)}_\alpha\rangle = \sum_i  c_{\alpha i}\int_T dt e^{-i[E_\alpha (t_0-t)+n \Omega t]} | \psi_i(t)\rangle.
\end{equation}
This requires storing $|\psi_i(t)\rangle$ at a relevant sampling rate during the propagation. We also note that evaluating Eq.~(\ref{eq:un}) amounts essentially to performing a discrete Fourier transform, which potentially provides routes to alternative efficient implementations to obtain $|u^{(n)}_\alpha\rangle$. 

\subsection{Applications}
In the following, we will demonstrate the method using four representative materials. For continuous-wave driving, we first consider graphene under circularly polarized light (CPL), a paradigmatic case in which a Floquet gap opens at the Dirac point\cite{oka_photovoltaic_2009}. This example illustrates why FBZ unfolding is essential for obtaining a clear Floquet band structure and for identifying hybridization gaps. We then turn to monolayer black phosphorus (BP), where we show that the Floquet eigenstates reconstructed from real-time propagation carry symmetry information of individual light-induced sidebands. Furthermore, we consider monolayer hexagonal boron nitride (hBN) driven by a finite CPL, demonstrating that the real-time analysis can be applied beyond the strictly periodic limit and can capture transient Floquet features under pulsed driving. Finally, we apply the same analysis to bulk tin sulfide (SnS), providing a 3D test case for the generality of the method.

\subsection{Graphene}\label{sec:graphene}
We first consider graphene under continuous-wave CPL driving, one of the paradigmatic examples of Floquet band engineering, where the breaking of time-reversal symmetry leads to the opening of a topological gap at the Dirac point\cite{haldane_model_1988,oka_photovoltaic_2009}. We compute the Floquet electronic structure along a momentum path crossing the K point using a CPL with photon energy $\Omega=4$~eV. Figure~\ref{fig:floquet_graphene} illustrates the resulting Floquet spectrum in three successive steps, highlighting the necessity of the unfolding procedure introduced above.

\begin{center}
\centering
\includegraphics[width=0.8\linewidth]{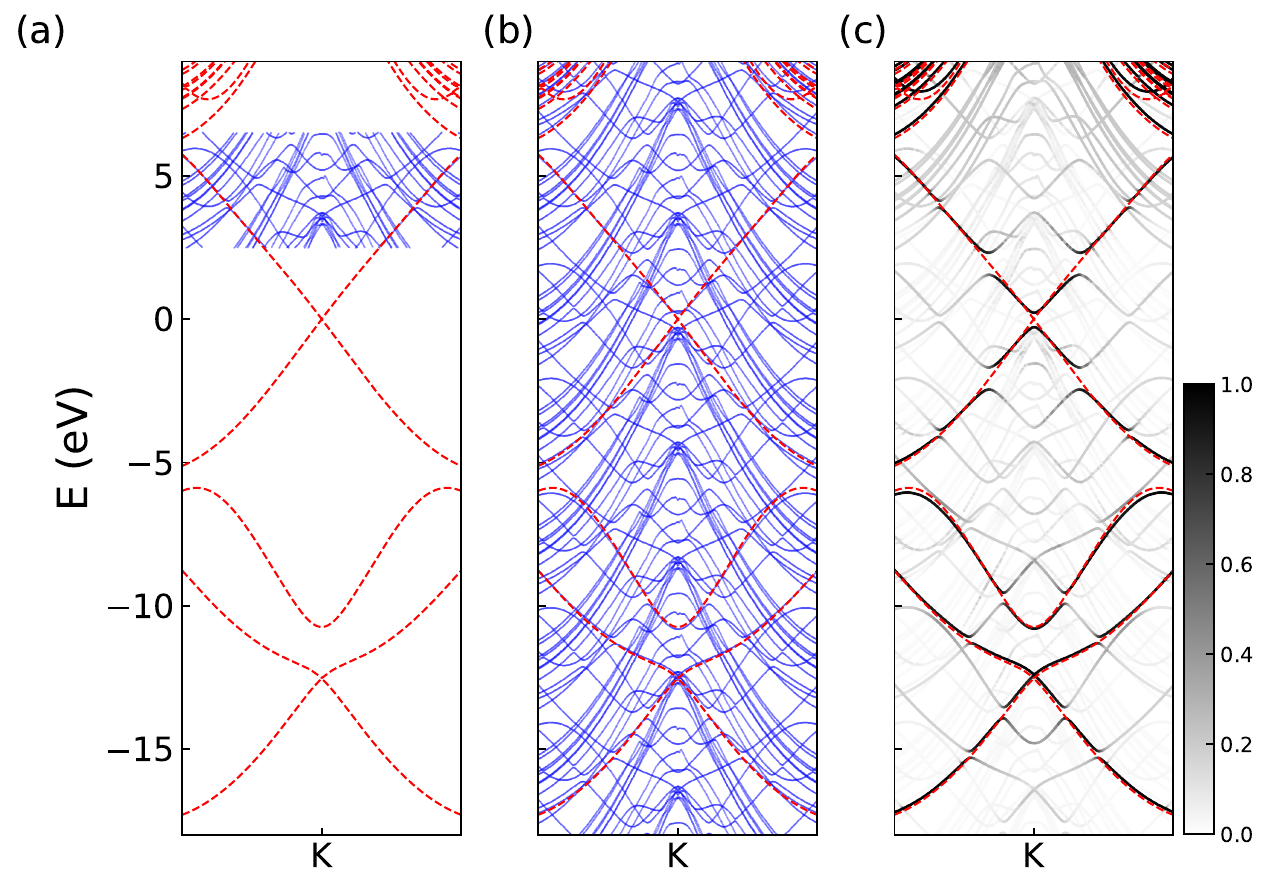}
\captionof{figure}{\textbf{Floquet band unfolding in graphene.} \textbf{a} Floquet band structure of graphene in the chosen Floquet Brillouin zone (FBZ), shown by blue lines. \textbf{b} Floquet band structure from all FBZs within the plotted energy window, shown by blue lines. \textbf{c} Unfolded Floquet spectrum in grayscale, with darker intensity indicating larger Floquet weight. The equilibrium band structure is plotted as red dashed lines in all panels.}
\label{fig:floquet_graphene}
\end{center}

In Fig.~\ref{fig:floquet_graphene}a, we show the equilibrium band structure together with the Floquet eigenvalues as they come out of the diagonalization of $U$ in Eq.~(\ref{eq:U_eig}). We see that, as expected, they are restricted to the displayed FBZ, centered around 4~eV. Since the Dirac point at K is set to 0~eV, one photon energy away from the chosen FBZ center, the physically relevant Floquet features are folded back into the chosen FBZ. As a consequence, the spectrum appears as a dense and largely unintelligible set of bands away from the region of interest. In particular, the characteristic Floquet gap opening at the Dirac point cannot be directly identified. This demonstrates that the raw quasi-energy spectrum in the chosen FBZ alone is insufficient for the interpretation of realistic material calculations.

To recover the full Floquet band structure, in Fig.~\ref{fig:floquet_graphene}b we unfold the spectrum by repeating the Floquet eigenvalues across all Floquet Brillouin zones (FBZs). While this restores the expected replica structure, the result is still difficult to interpret. The repeated replicas generate a large number of overlapping bands throughout the entire energy window and, especially around the K point, several folded bands intersect and hybridize simultaneously. Although the Floquet-induced avoided crossings are now formally present, the physically relevant hybridization gaps remain obscured by the large number of replica bands.

Finally, Fig.~\ref{fig:floquet_graphene}c shows the unfolded Floquet spectrum weighted by the norm of the zeroth Floquet harmonic in Eq.~(\ref{eq:weight}). Here the usefulness of the unfolding procedure combined with the Floquet weights becomes evident. The irrelevant replica features are strongly suppressed, while the states with dominant equilibrium character remain clearly visible. As a result, the Floquet-engineered electronic structure becomes directly interpretable: a clear topological gap opens at the Dirac point at K, and additional hybridization gaps appear whenever equilibrium bands come into resonance with Floquet replicas shifted by the driving frequency. The weighting procedure therefore exposes the physically important parts of the Floquet spectrum while retaining the full information about the replica structure and hybridization processes.

\subsection{Black phosphorus}
\label{sec:bp}

Recently, BP has emerged as a promising platform for Floquet engineering. Most existing pump-probe experiments have focused on bulk BP, whose puckered layered structure gives rise to a pronounced in-plane anisotropy between the armchair (AC) and zigzag (ZZ) directions\cite{jung2020black}. This anisotropy leads to a strong polarization dependence of the Floquet response. In particular, AC-polarized pumping can induce pronounced Floquet band renormalization, whereas ZZ-polarized pumping mainly generates light-induced replica bands without sizable hybridization gaps\cite{zhou2023pseudospin,zhou2023Floquet,bao2024light}. When different probe polarizations are considered, the spectral weights are further redistributed among different light-induced sidebands in a pump-probe-geometry-dependent manner\cite{SYZhou2024spot}. This behavior is governed by Floquet optical selection rules\cite{fan2025floquet}, and can be further controlled by tilting the crystal orientation\cite{bao2025floquet}. Although these experimental observations mainly concern bulk BP, the essential in-plane anisotropy and relevant symmetry properties are retained in monolayer BP. Therefore, here we use monolayer BP, shown in the inset of Fig.~\ref{fig:floquet_BP}a, as a minimal and symmetry-transparent representative system to demonstrate the validity and applicability of our approach.

We first apply our method to monolayer BP and examine its Floquet band renormalization. As shown in Fig.~\ref{fig:floquet_BP}a, we consider an AC-polarized continuous-wave driving, with the photon energy set to the band-gap energy. Consistent with previous studies\cite{zhou2023pseudospin,zhou2023Floquet}, pronounced band renormalization emerges between different Floquet sidebands, indicating symmetry-allowed hybridization under AC pumping. Beyond the quasi-energy spectrum, our approach can also resolve the symmetry character of individual light-induced sidebands at a given $\mathbf{k}$ point. In equilibrium, at the $\Gamma$ point, the valence band (VB) and conduction band (CB) have odd and even glide-mirror symmetries under $\widetilde{\mathcal{M}}_{\rm AC}$, respectively. Under AC pumping, the Floquet optical selection rules\cite{fan2025floquet} show that the $n=0$ Floquet sidebands retain the symmetry of their parent equilibrium bands. Therefore, the $n=0$ Floquet sideband for the VB (CB), namely, the $n=0$ Floquet VB (CB), remains odd (even). In contrast, upon absorption or emission of one pump photon, the glide-mirror symmetry is reversed. As a result, the $n=1$ Floquet VB becomes even, while the $n=-1$ Floquet CB becomes odd under $\widetilde{\mathcal{M}}_{\rm AC}$.

\begin{center}
\centering
\includegraphics[width=0.65\linewidth]{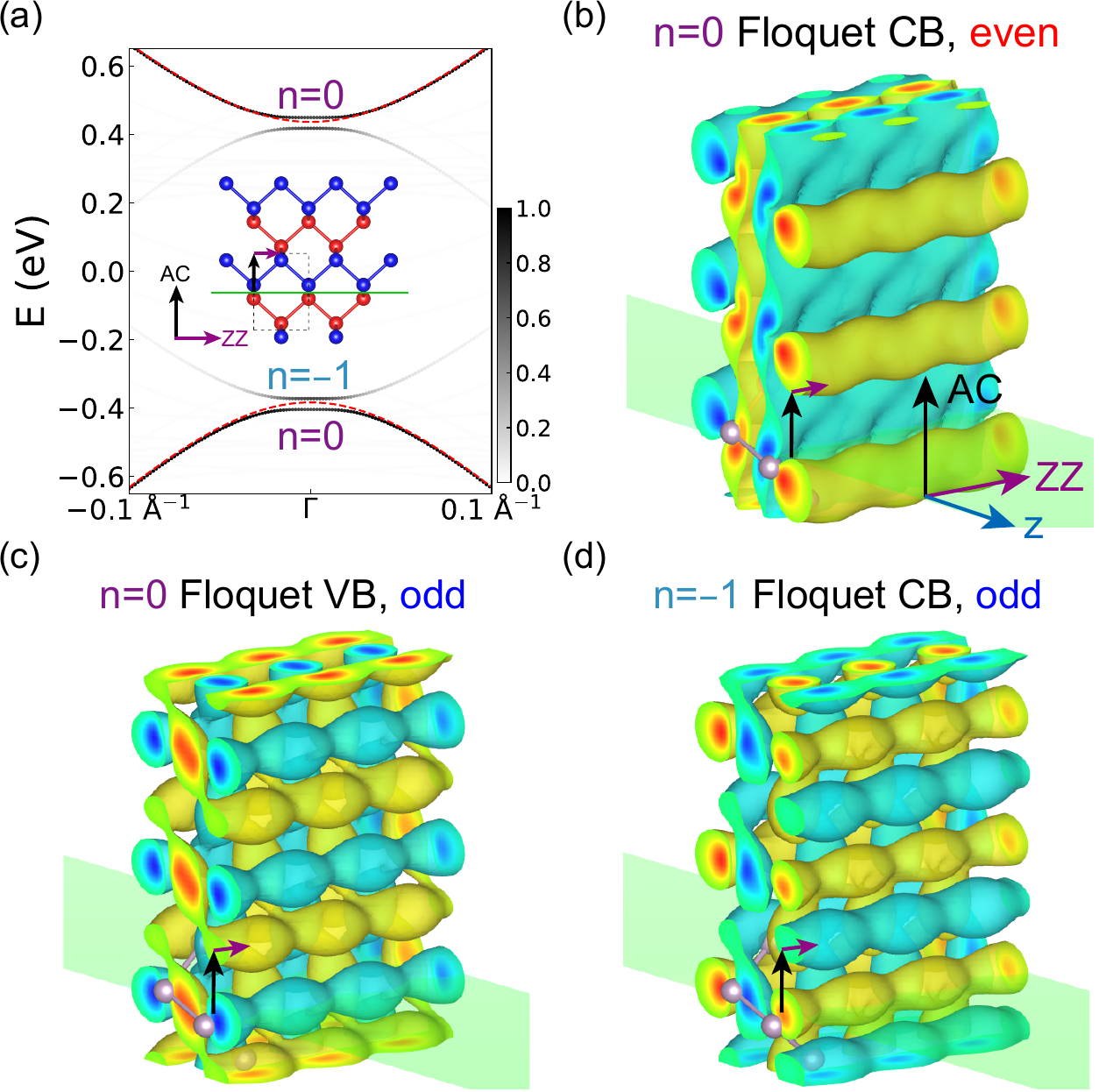}
\captionof{figure}{\textbf{Real-space symmetry of Floquet sidebands in monolayer black phosphorus.} \textbf{a} Equilibrium band structure of monolayer black phosphorus (BP) and the corresponding Floquet band structure obtained under resonant armchair (AC) pumping with photon energy $\Omega=0.82$~eV, both plotted along the AC direction. The equilibrium bands are shown by red dashed lines, while the Floquet bands are shown in grayscale. The inset shows the lattice structure of monolayer BP, where blue and red atoms denote the upper and lower sublayers, respectively. The black dashed lines denote the unit cell. Monolayer BP has a glide-mirror symmetry $\widetilde{\mathcal{M}}_{\rm AC}$, which consists of a mirror operation $\mathcal{M}_{\rm AC}$ with respect to the green plane perpendicular to the AC direction, followed by translations by half a lattice vector along both the AC and zigzag (ZZ) directions, as indicated by the black and purple arrows at the edge of the unit cell. Darker grayscale intensity indicates larger Floquet weight, and $n$ denotes the Floquet index. \textbf{b} Real part of the real-space wavefunction of the $n=0$ Floquet conduction band (CB) at the $\Gamma$ point in \textbf{a}, which is even with respect to the glide-mirror symmetry $\widetilde{\mathcal{M}}_{\rm AC}$. The yellow and blue isosurfaces denote the positive and negative lobes of the real-space wavefunction, respectively. \textbf{c} Same as \textbf{b}, but for the $n=0$ Floquet valence band (VB) with odd glide-mirror symmetry. \textbf{d} Same as \textbf{b}, but for the $n=-1$ Floquet CB, which also has odd symmetry.}
\label{fig:floquet_BP}
\end{center}

These symmetry assignments can be directly verified from the real-space Floquet wavefunctions obtained in our calculation via Eq.~(\ref{eq:un}). Here we visualize the real part of the Floquet wavefunction for various Floquet bands at the $\Gamma$ point for the driving condition used in Fig.~\ref{fig:floquet_BP}a. Figure~\ref{fig:floquet_BP}b shows the real part of the real-space wavefunction of the $n=0$ Floquet CB. After applying the mirror operation $\mathcal{M}_{\rm AC}$ with respect to the green plane, followed by the half-lattice translations along the AC and ZZ directions indicated by the black and purple arrows, the yellow isosurfaces are mapped onto yellow isosurfaces in the same sublayer. This confirms that the $n=0$ Floquet CB is even under the glide-mirror operation $\widetilde{\mathcal{M}}_{\rm AC}$. By contrast, for the $n=0$ Floquet VB shown in Fig.~\ref{fig:floquet_BP}c, the same glide-mirror operation maps blue isosurfaces onto yellow isosurfaces, indicating a sign reversal of the wavefunction and hence odd symmetry. The same analysis applies to the $n=-1$ Floquet CB in Fig.~\ref{fig:floquet_BP}d, which is also odd under $\widetilde{\mathcal{M}}_{\rm AC}$. Therefore, the $n=0$ Floquet VB and the $n=-1$ Floquet CB have the same glide-mirror symmetry at the $\Gamma$ point. Their hybridization is symmetry allowed, leading to the Floquet band repulsion observed in Fig.~\ref{fig:floquet_BP}a. This example demonstrates that our method not only reproduces the pump-polarization-dependent Floquet band structures of BP, but also captures the symmetry character of individual Floquet sidebands at the wavefunction level. It therefore provides a tool for the full analysis of Floquet features, including symmetry, matrix elements and selection rules.

\subsection{Hexagonal boron nitride}\label{sec:hbn}

Although the construction of the Floquet states formally assumes time periodicity, the method can be extended to situations where the Hamiltonian is only approximately periodic over a finite time window, such as in the presence of pulsed driving fields. In this case, the overlaps $\langle \psi_i(t) | \psi_j(t+T) \rangle$ define an effective evolution operator that captures the instantaneous Floquet properties of the system. We emphasize that the applicability of this time-local Floquet interpretation is controlled by the variation of the complete self-consistent Hamiltonian over the analysis window, rather than by the field strength alone. For a strictly periodic Hamiltonian, the construction remains valid also in the strong-driving regime. To demonstrate this point, in this section we present the Floquet band structure of monolayer hBN beyond the strictly periodic limit. As shown in the upper panels of Fig.~\ref{fig:floquet_hbn_time}, the driving field is a CPL pulse rather than a continuous wave, with a time-dependent intensity governed by an envelope function. The pulse has a resonant carrier photon energy of $\Omega=4.139$ eV and a finite duration of $T_{\mathrm{pu}}=9T$, where $T=2\pi/\Omega$ is the optical quasi-period of the laser field. Similar to the former cases, we calculate the Floquet bands from the overlap matrix $\langle \psi_i(t) | \psi_j(t+T) \rangle$ along the momentum path $\text{-}\mathrm{K}-\Gamma-\mathrm{K}$. The key distinction here is that the overlap matrix no longer possesses time-translational symmetry. Therefore, instead of fixing the starting time $t$ in the Floquet analysis, it is instructive to examine how the resulting Floquet band structure depends on the choice of starting time.

We therefore choose three representative starting times, $t=T$, $4T$, and $7T$, to perform Floquet analyses during the second, fifth, and eighth optical cycles of the pulse, respectively. As shown in Fig.~\ref{fig:floquet_hbn_time}a, when the analysis window is centered near the beginning of the pulse, the Floquet weights of the replica bands are small except for the $n=0$ Floquet sidebands, which essentially coincide with the equilibrium band structure. By contrast, when the analysis window is placed around the peak of the pulse, as shown in Fig.~\ref{fig:floquet_hbn_time}b, clear replicas of the VB and CB emerge. In addition, since the pump photon energy is resonant with the direct gap, the first Floquet replicas of the valence band maximum (VBM) and conduction band minimum (CBM) are brought close to the original CBM and VBM, respectively. Pronounced band renormalization then appears near the VBM and CBM around the \text{-}K valley under the left-CPL, as indicated by the blue arrows. By contrast, the CB and VB around the K valley show no substantial renormalization. This valley-dependent Floquet band renormalization is closely related to the optical selection rule rooted in the valley symmetry of monolayer hBN\cite{Cao_2012,zhang2022hbn}, as discussed below. In Fig.~\ref{fig:floquet_hbn_time}c, where the Floquet analysis starts at $t=7T$, close to the end of the pulse, the Floquet replica bands fade again, leaving mainly the $n=0$ Floquet sidebands. The results in Figs.~\ref{fig:floquet_hbn_time}a and c indicate that the effects of Floquet sidebands and band renormalization become negligible when the field is weak, as expected. These observations are consistent with the fact that Floquet band structures can remain meaningful in transient regimes where strict periodicity is not fulfilled, provided that the system evolves slowly on the timescale of the driving period. The present approach therefore provides a practical route for performing Floquet analysis directly within real-time simulations of non-equilibrium systems.

We now analyze the origin of the valley-dependent Floquet response in monolayer hBN. The band-edge states at K and \text{-}K can be characterized by discrete angular momenta associated with $C_3$ symmetry\cite{zhang2022hbn}.
The relevant optical matrix element induced by CPL is $\langle c\mathbf{k}|(\hat{p}_x\pm i\hat{p}_y)|v\mathbf{k}\rangle$,
where $\mathbf{k}=$ K or \text{-}K, and $|c\mathbf{k}\rangle$ and $|v\mathbf{k}\rangle$ denote the wavefunctions of CB and VB at momentum $\mathbf{k}$, respectively. The two operators $\hat{p}_{\pm}=\hat{p}_x\pm i\hat{p}_y$ correspond to opposite circular polarizations. Under a $C_3$ rotation, the Bloch states transform as $\hat{C}_3|c\mathbf{k}\rangle=e^{i2\pi m_c(\mathbf{k})/3}|c\mathbf{k}\rangle$
and
$\hat{C}_3|v\mathbf{k}\rangle=e^{i2\pi m_v(\mathbf{k})/3}|v\mathbf{k}\rangle$,
where the discrete angular momenta satisfy $m_{c(v)}(\mathbf{k})=0,\pm1~(\mathrm{mod}~3)$. The circularly polarized momentum operators transform under $C_3$ as
$\hat{C}_3\hat{p}_{\pm}\hat{C}_3^{-1}=e^{i2\pi m_{\pm}/3}\hat{p}_{\pm}$,
indicating that a circularly polarized photon carries angular momentum $m_{\pm}=\pm1$. The optical transition matrix element is therefore nonzero only when the discrete angular momentum is conserved, namely
$m_c(\mathbf{k})-m_v(\mathbf{k})=m_{\pm}$. Time-reversal symmetry relates the two valleys and complex-conjugates their $C_3$ eigenvalues, so the corresponding angular momenta are opposite modulo three. Therefore, the allowed circular polarization is reversed between K and \text{-}K valleys. 

This selection rule predicts that reversing the pump helicity should interchange the valley with the stronger Floquet band renormalization. We therefore performed an additional calculation using a right-CPL under the same conditions as in Fig.~\ref{fig:floquet_hbn_time}b. As indicated by the purple arrows in Fig.~\ref{fig:floquet_hbn_time}d, the CB and VB near the K point exhibit strong band renormalization, whereas those around the \text{-}K point remain almost unchanged. This valley-asymmetric band renormalization, together with its reversal under opposite pump helicity, validates the implementation and demonstrates that the real-time Floquet analysis can capture valley-selective light-matter coupling and nonequilibrium Floquet band engineering in realistic materials.

\begin{center}
\centering
\includegraphics[width=\linewidth]{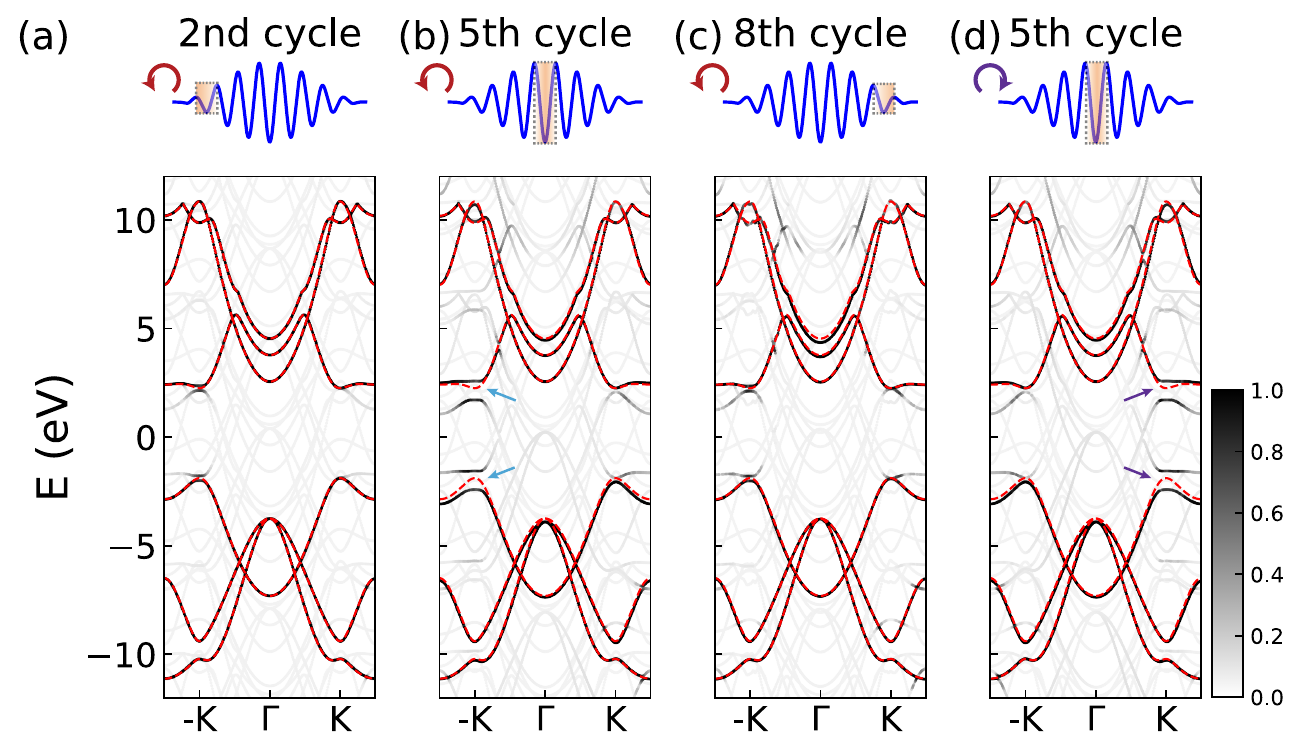}
\captionof{figure}{\textbf{Transient Floquet band structure of monolayer hBN under a finite circularly polarized pulse.} \textbf{a} Floquet band structure obtained from an analysis window taken within the second cycle of the pump, with starting time $t=T$, as indicated by the dashed black box in the blue pulse profile. The red circular arrow denotes left-circular polarization. The Floquet band structure is plotted in grayscale, while the equilibrium band structure is overlaid as red dashed lines. \textbf{b} Same as \textbf{a}, but with the analysis window taken within the fifth cycle of the left-circularly polarized light (CPL) pump, starting at $t=4T$. The blue arrows indicate band renormalization near the \text{-}K point. \textbf{c} Same as \textbf{a}, but with the analysis window taken within the eighth cycle of the left-CPL pump, starting at $t=7T$. \textbf{d} Same as \textbf{b}, but for a right-CPL pump, indicated by the purple circular arrow. The purple arrows indicate band renormalization near the K point.}
\label{fig:floquet_hbn_time}
\end{center}

\subsection{Bulk tin sulfide}
\label{sec:bulk_sns}

Having demonstrated the method for 2D systems, we next apply the real-time Floquet analysis to bulk SnS, an anisotropic layered semiconductor shown in the inset of Fig.~\ref{fig:sns}a\cite{Rodin2016-ie,Lin2018-do}, to test its applicability to a fully periodic 3D solid. We calculate the Floquet band structure along the in-plane 
$\mathrm{X}$-$\Gamma$-$\mathrm{Y}$ momentum path under linearly polarized continuous-wave driving with a near-resonant photon energy. Two pump geometries are considered, which we refer to as AC and ZZ pumps. The detailed polarization geometry is 
specified in Methods.

As shown in Fig.~\ref{fig:sns}a, the Floquet band structure under AC-polarized driving exhibits pronounced band renormalization near the $\Gamma$X valley, as indicated by the blue arrows, whereas the bands near the $\Gamma$Y valley remain much less affected. This valley-selective renormalization is consistent with previous symmetry-based analyses of multivalley SnS\cite{fragkos2026symmetry}, where the polarization-dependent Floquet hybridization is associated with the symmetry compatibility between $n=0$ Floquet bands and the relevant photon-shifted replicas\cite{Claassen:2016ge}.

In contrast, the ZZ pump geometry shows a complementary polarization dependence in Fig.~\ref{fig:sns}b. In this case, stronger hybridization features appear near the $\Gamma$Y valley at crossings between the $n=0$ Floquet bands and their replicas, as indicated by the purple arrows. Near the $\Gamma$X valley, however, the Floquet bands closely follow the equilibrium bands, and no appreciable renormalization is observed. These polarization-dependent features are consistent with previous symmetry-guided analyses of driven SnS\cite{fragkos2026symmetry}. The results demonstrate that the present real-time Floquet analysis is not restricted to 2D slab geometries, but can also be applied directly to fully periodic 3D bulk materials.

\begin{center}
\centering
\includegraphics[width=0.7\linewidth]{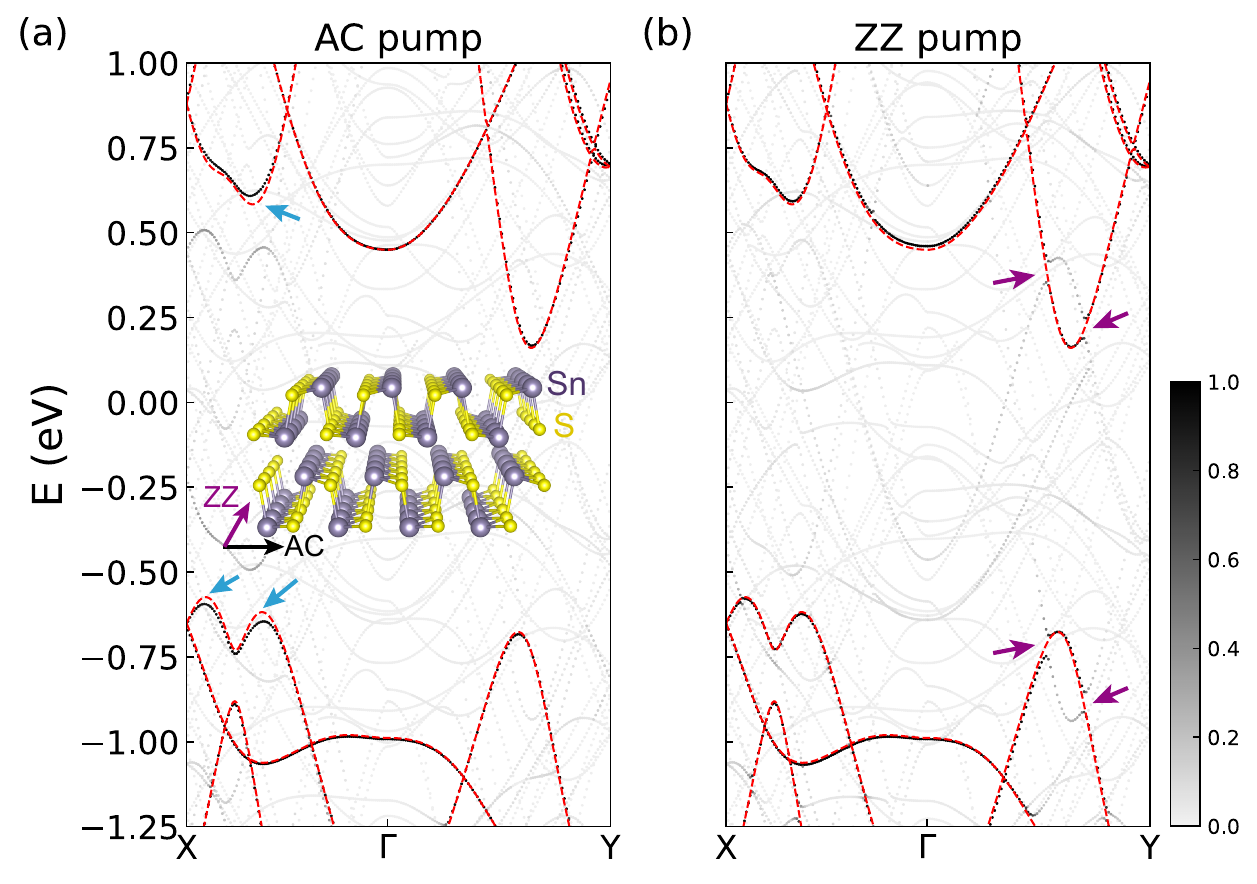}
\captionof{figure}{\textbf{Polarization-dependent Floquet band renormalization in bulk SnS.} \textbf{a} Grayscale Floquet band structure of bulk SnS along the in-plane momentum path $\mathrm{X}-\Gamma-\mathrm{Y}$, calculated under an AC-polarized cosine driving field. The reduced coordinates of X and Y are $(0.5,0,0)$ and $(0,0.5,0)$, respectively, with $\mathrm{X}-\Gamma$ along the AC direction and $\Gamma-\mathrm{Y}$ along the ZZ direction. The inset shows the crystal structure of bulk SnS. Purple and yellow spheres denote Sn and S atoms, respectively. The blue arrows indicate band renormalization near the $\Gamma X$ valley. \textbf{b} Same as \textbf{a}, but for the ZZ pump geometry. The purple arrows indicate the corresponding hybridization features near the $\Gamma Y$ valley. In both panels, the equilibrium band structure is overlaid as red dashed lines.}
\label{fig:sns}
\end{center}

\section*{Discussion}
\label{sec:discussion}

We have introduced a real-time TDDFT approach for extracting Floquet quasi-energies and Floquet states from the one-period evolution operator, thereby avoiding the explicit construction and diagonalization of the Floquet Hamiltonian. In addition, we developed an unfolding procedure based on the harmonic decomposition of Floquet states. Within this framework, the zeroth-harmonic component provides a physically transparent measure of equilibrium band character, enabling the identification of relevant non-equilibrium Floquet branches beyond the reduced-zone representation. The reconstructed Floquet states also retain real-space and symmetry information, enabling analysis of light-induced sidebands and their selection rules.

We demonstrated the method in four representative 2D and 3D materials, showing its applicability to realistic first-principles simulations. Beyond its computational advantages, the present work establishes a broader conceptual perspective on Floquet theory. Rather than treating Floquet theory solely as a formal description of strictly periodic steady states, our results show that Floquet-like electronic structures can be meaningfully identified in transient driving regimes, where local light-dressed sidebands emerge within a few optical cycles. This supports the use of Floquet analysis for non-equilibrium electronic structure even under finite pulses and incomplete time periodicity. At stronger excitation, the effect of carrier excitation or changes in the screening may lead to appreciable evolution of the self-consistent Hamiltonian over successive cycles. In this regime, the Floquet interpretation of the eigenstates of $U(t+T,t)$ becomes less well defined. Nevertheless, the present framework provides a means of diagnosing this breakdown: the extracted quasienergies and Floquet states can be compared between consecutive cycles and, more importantly, with observables obtained directly from the \textit{same} real-time simulation, such as time- and angle-resolved photoemission spectroscopy. A loss of cycle-to-cycle stability or a mismatch between the reconstructed Floquet bands and the observed photoemission features would signal the breakdown of a quasistationary Floquet description.

Looking forward, the present approach could be extended to a broad class of driven quantum systems and non-equilibrium phenomena, including correlated materials, collective excitations, and coupled electron-phonon dynamics. The evolution-operator perspective also suggests a natural extension beyond a single optical cycle. By constructing and comparing multi-cycle evolution operators, one could quantify how Floquet-like quasiparticles build up, persist and decay during finite pulses.
More broadly, this framework opens a route to identifying and interpreting light-dressed states in realistic materials under ultrafast driving conditions. Finally, we stress that while we present the method here within a TDDFT framework it can readily be applied to any real-time propagation scheme that evaluates the time-dependent wavefunctions or even extended in spirit to density-matrix propagation schemes.

\begin{methods}
\label{sec:methods}
\subsection{Floquet gauge redundancy and the interpretation of replica weights}
\label{app:floquet_redundancy}

The Floquet representation contains a gauge redundancy associated with the definition of the quasi-energy modulo the driving frequency. Starting from Eqs.~(\ref{eq:F_def}) and (\ref{eq:phi_def}), one obtains
\begin{equation}
\begin{aligned}
|\Psi_\alpha(t)\rangle
&= e^{-iE_\alpha t}\sum_n e^{in\Omega t}|u_\alpha^{(n)}\rangle \\
&= e^{-i(E_\alpha-r\Omega)t}\sum_n e^{i(n-r)\Omega t}|u_\alpha^{(n)}\rangle \\
&= e^{-iE_{\alpha^\prime} t}\sum_{n'} e^{in'\Omega t}|u_\alpha^{(n'+r)}\rangle \\
&= e^{-iE_{\alpha^\prime} t}\sum_n e^{in\Omega t}|u_{\alpha^\prime}^{(n)}\rangle ,
\end{aligned}
\label{eq:redundancy}
\end{equation}
where we have defined $E_{\alpha^\prime}=E_\alpha-r\Omega$ and
$|u_{\alpha^\prime}^{(n)}\rangle = |u_\alpha^{(n+r)}\rangle$.
This shows that shifting the quasi-energy by an integer multiple of $\Omega$ is accompanied by a corresponding shift of the harmonic components. The transformation does not change the physical time-dependent Floquet state, but only changes the Floquet gauge used to represent it\cite{rudner2020floquet}.

This redundancy is schematically illustrated in Fig.~\ref{fig:floquet_redundancy}. The left column denotes the representation in the reduced FBZ, centered at $E_\alpha$, while the right column denotes the representation in the $r=-1$ extended-zone copy, corresponding to the quasi-energy $E_{\alpha^\prime}=E_\alpha+\Omega$. In this gauge, the zeroth harmonic component is not the original $|u_\alpha^{(0)}\rangle$, but rather
\begin{equation}
|u_{\alpha^\prime}^{(0)}\rangle = |u_\alpha^{(-1)}\rangle .
\end{equation}

This relation is useful for understanding how the unfolding weights are evaluated. In the reduced FBZ, the equilibrium-character weight is defined in Eq.~(\ref{eq:weight}) as $w_\alpha=| u_\alpha^{(0)}|$.
For an extended-zone copy labeled by $r$, the same definition is applied in the corresponding Floquet gauge,
\begin{equation}
w_{\alpha}
=|u_{\alpha^\prime}^{(0)}|=| u_\alpha^{(r)}|,
\end{equation}
where we have used $|u_{\alpha^\prime}^{(n)}\rangle = |u_\alpha^{(n+r)}\rangle$ in Eq.~(\ref{eq:redundancy}).
For example, for the $r=-1$ extended-zone copy, which corresponds to the $+\Omega$ energy branch, the weight is
\begin{equation}
w_{\alpha}
=
| u_\alpha^{(-1)}|.
\end{equation}
This is the quantity used to assign spectral weight to the $E_\alpha+\Omega$ replica in the unfolded Floquet band structure in the code.

\begin{center}
\centering
\includegraphics[width=0.5\linewidth]{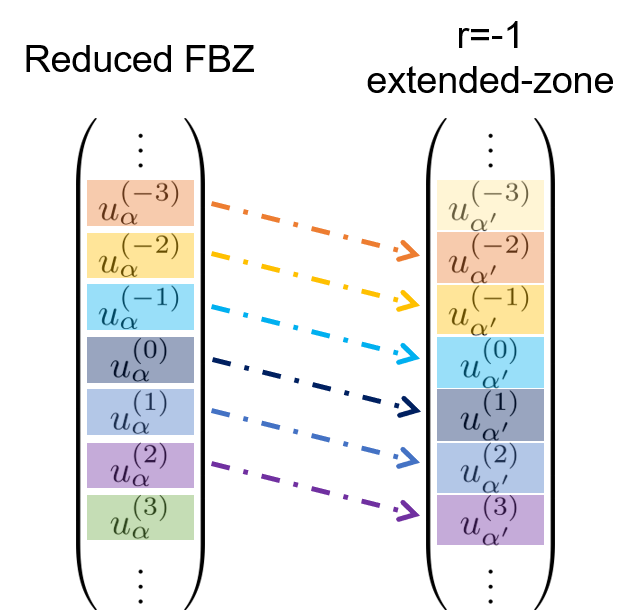}
\captionof{figure}{\textbf{Schematic illustration of the Floquet gauge redundancy.} The left column shows the harmonic components $u_\alpha^{(n)}$ in the reduced FBZ representation with quasi-energy $E_\alpha$. The right column shows the equivalent representation in the $r=-1$ extended-zone copy as an example, where the quasi-energy is shifted to $E_{\alpha^\prime}=E_\alpha+\Omega$. Under this transformation, the harmonic components are relabeled according to $u_{\alpha^\prime}^{(n)}=u_\alpha^{(n-1)}$. In particular, the zeroth harmonic in the $r=-1$ representation is $u_{\alpha^\prime}^{(0)}=u_\alpha^{(-1)}$.}
\label{fig:floquet_redundancy}
\end{center}

\subsection{Computational details}
\label{app:computational_details}

All calculations were performed within real-time TDDFT as implemented in the {\tt Octopus} code\cite{andrade2015real,tancogne2020octopus}. We employed norm-conserving Hartwigsen-Goedecker-Hutter pseudopotentials\cite{hartwigsen1998relativistic} within the local density approximation and represented the electronic wave functions on a real-space grid with a spacing of 0.36~$a_0$ in all calculations. Spin-orbit coupling was not included in the present analysis.

The graphene monolayer was modeled using periodic boundary conditions in two dimensions with a vacuum spacing of 80~$a_0$ in the out-of-plane direction. Brillouin-zone (BZ) sampling was performed using a $6\times6\times1$ $k$-point grid, while the Floquet bands were evaluated along a momentum path crossing the K point. The system was driven by a continuous-wave CPL with photon energy $\Omega=4$~eV and peak field strength of approximately 2.3~MV/cm. Time propagation was carried out with a time step of 0.08~a.u., and the Floquet analysis was performed directly from the real-time propagation using the one-period evolution operator. Floquet weights and harmonic components up to order $n=\pm8$ were included in the analysis.

For monolayer BP, the KS equations were solved self-consistently with a relative density convergence threshold of $10^{-8}$. A $\Gamma$-centered $12\times10\times1$ k-point grid was used for BZ sampling. The out-of-plane $z$ direction, normal to the monolayer plane, was treated with non-periodic boundary conditions using a simulation box length of 120~$a_0$. The lattice parameters along the AC and ZZ directions were 4.625~\AA{} and 3.299~\AA{}, respectively. The calculated band gap of monolayer BP was 0.82~eV. For the Floquet analysis, the system was driven by a linearly polarized cosine field along the AC direction, with a pump peak intensity of $I_{\rm pu}=4\times10^8$~W/cm$^2$. To ensure convergence, the time step was set to 0.07~a.u. ($1.69\times10^{-3}$~fs). Floquet weights and harmonic components up to order $n=\pm20$ were included in the analysis.

The hBN monolayer was treated using a 2D periodic geometry, with periodic boundary conditions applied in the in-plane directions and a nonperiodic boundary condition along the out-of-plane direction. A vacuum region of $40~a_0$ was included to avoid spurious interactions between periodic images. The BZ was sampled using a $6\times6\times1$ k-point mesh. The Floquet electronic structure was analyzed under a CPL pulse containing 9 optical cycles, with a photon energy of $\Omega=4.139$ eV and a peak field strength of 0.58~MV/cm. The time-dependent simulation was conducted with a time step of 0.08 a.u. ($1.94\times10^{-3}$ fs). The Floquet states were obtained directly from the real-time propagation framework, with harmonic sideband components included up to order $n=\pm20$.

For bulk SnS, the in-plane lattice constants along the AC and ZZ directions were set to 4.421~\text{\AA} and 3.989~\text{\AA}, respectively, and the out-of-plane lattice constant was set to 11.174~\text{\AA}. Brillouin-zone sampling was performed using a $\Gamma$-centered $10\times12\times4$ k-point grid. The self-consistent KS calculation was converged to a relative density threshold of $10^{-8}$, yielding an indirect band gap of 0.95~eV. The Floquet analysis was performed under linearly polarized cosine fields with photon energy $\Omega=1.1$~eV and peak intensity $1.0\times10^{10}~\text{W}/\text{cm}^{2}$. Two driving geometries were considered. For the AC pump, the field was polarized entirely in plane along the AC direction. For the ZZ pump, the in-plane projection of the field was along the ZZ direction, while the polarization vector was tilted by $45^{\circ}$ out of the layer plane, thereby including an out-of-plane component. A time step of 0.07~a.u. ($1.69\times10^{-3}$~fs) was used to ensure numerical convergence. Harmonic sideband components up to order $n=\pm40$ were retained in the analysis.
\end{methods}

\begin{addendum}
\item[Data availability]
~\\
The datasets generated and/or analyzed during the current study are not publicly available due to the large size of the raw real-time TDDFT datasets, but are available from the corresponding authors on reasonable request.

\item[Code availability]
~\\
{\tt Octopus} code that was used for performing all of the simulations in the main text is available at \\
\href{https://gitlab.com/octopus-code/octopus}{https://gitlab.com/octopus-code/octopus}.

\item[Acknowledgements]
~\\
We acknowledge useful discussions with Ofer Neufeld and Samuel Beaulieu.

\item[Funding]
~\\
This work was supported by the Cluster of Excellence 'CUI: Advanced Imaging of Matter'- EXC 2056 - project ID 390715994, SFB-925 "Light induced dynamics and control of correlated quantum systems" - project ID 170620586  of the Deutsche Forschungsgemeinschaft (DFG), the European Research Council (ERC-2024-SyG-UnMySt - 101167294), the European Union Marie Sklodowska-Curie Doctoral Networks (TIMES, Grant No. 101118915; and SPARKLE, Grant No. 101169225), the Max Planck-New York City Center for Non-Equilibrium Quantum Phenomena and the Italian Ministry of University and Research (MUR) under the PRIN 2022738 (Grant No. 2022PX279E 003). The Flatiron Institute is a division of the Simons Foundation. The funders played no role in study design, data analysis and interpretation, or the writing of this manuscript.
\end{addendum}

\subsection{References}
\bibliography{main}

\begin{addendum}
\item[Author contributions]
~\\
A.R. and H.H. conceived the project. H.H., R.L., and B.F. developed and tested the Floquet-analysis code. R.L. and B.F. performed the first-principles calculations and prepared all figures. B.F., R.L., H.H., and U.D.G. analyzed and interpreted the theoretical results. B.F., R.L., and H.H. wrote the manuscript with input from all authors.

\item[Competing interests]
~\\
The authors declare no competing financial or non-financial interests.
\end{addendum}

\end{document}